\documentclass [12pt]{article}
\begin{document}

\title{\textbf{Trion ladder diagrams}}
\author{M. Combescot and O. Betbeder-Matibet
 \\ \small{\textit{GPS, Universit\'e Denis Diderot
and Universit\'e Pierre et Marie Curie,
CNRS,}}\\ \small{\textit{Tour 23, 2 place Jussieu, 75251
Paris Cedex 05, France}}}
\date{}
\maketitle

\begin{abstract}

We first derive a new ``commutation technique'' for an exciton
interacting with electrons, inspired from the one we recently
developed for excitons interacting with excitons. These
techniques allow to take \emph{exactly} into account
the possible exchanges between carriers. We use it to get the
$\mathrm{X}^-$ trion creation operator in terms of exciton and
free-electron creation operators. In a last part we  generate the
ladder diagrams associated to these trions. Although similar to
the exciton ladder diagrams, with the hole replaced by an
exciton, they are actually much more tricky : (i) Due to the
composite nature of the exciton, one cannot identify an
exciton-electron potential similar to the Coulomb potential
between electron and hole ; (ii)  the spins are unimportant for
excitons while they are crucial for trions, singlet and triplet
states having different energies.
\end{abstract}

\vspace{2cm}


\newpage

The stability of semiconductor trions has been predicted long
ago $^{(1,2)}$. However, their binding energies being extremely
small in bulk materials, clear experimental evidences $^{(3,4)}$
of these exciton-electron bound states have been achieved
recently only, due to the development of good semiconductor
quantum wells, the reduction of dimensionality enhancing all
binding energies.

It is now possible to study these exciton-electron bound
states as well as their interactions $^{(5-7)}$ with other
carriers. However many-body effects with trions are even more
subtle than many-body effects with excitons, due to their
composite nature : Being made of indiscernable fermions, the
interchange of these fermions with other carriers are quite
tricky to handle properly.

We have recently developed a novel procedure to treat many-body
effects between close-to-boson particles $^{(8)}$, such as
excitons in semiconductors. It allows to take \emph{exactly} into
account the possible exchanges between their components. In
this paper, we first generate a similar procedure for an exciton
interacting with electrons. We clearly see appearing an
exciton-electron coupling induced by Coulomb interaction
\emph{and} an exciton-electron coupling induced by the
possible exchange of the exciton electron with the other
electrons. .

We use this commutation technique to get the trion
creation operators in terms of exciton and free-electron
creation operators and we show how, in the case of trions, all
exchange couplings can be cleverly hidden in the prefactors of
the trion operators, provided that  these prefactors have a very
specific invariance.

In a last part, we derive the trion ladder diagrams between
a``free'' exciton and a free electron. They are conceptually
similar to the exciton ladder diagrams $^{(9)}$ between a free
electron-hole pair, except for some quite specific difficulties
associated to the composite nature of the exciton.

\section{Commutation technique for an exciton
interacting with electrons}

The commutation technique for an exciton
interacting with electrons relies on two parameters
$\Xi_{n'\mathbf{k'}n\mathbf{k}}^{\mathrm{dir}}$ and
$\Lambda_{n'\mathbf{k'}n\mathbf{k}}$ which are such that
\begin{equation}
\left[V_{n,\sigma_n}^\dag\,
,a_{\mathbf{k},s}^\dag
\right]=
\sum_{n',\mathbf{k}'}\Xi_{n'\mathbf{k}'n
\mathbf{k}}
^{\mathrm{dir}}\, 
B_{n',\sigma_n}^\dag\,a_{\mathbf{k'},s}^\dag\
,
\end{equation}
\begin{equation}
\left[D_{n',\sigma_{n'};n,\sigma_n}\, ,
a_{\mathbf{k},s}^\dag\right]=
\sum_{\mathbf{k}'}\delta_{\sigma_{n'},s}\,
\Lambda_{n'\mathbf{k}'n\mathbf{k}}
\,a_{\mathbf{k'},\sigma_n}^\dag\ ,
\end{equation}
the ``creation potential'' $V_{n,\sigma_n}^\dag$ and the
deviation-from-boson operator
$D_{n',\sigma_{n'};n,\sigma_n}$ being defined as for excitons
interacting with excitons
$^{(8)}$ by $\left[H,B_{n,\sigma_n}^\dag\right]=E_n
B_{n,\sigma_n}^\dag+V_{n,\sigma_n}^\dag$ and $\left[B_{n',
\sigma_{n'}},B_{n,\sigma_n}^\dag\right]=\delta_{n,n'}\,
\delta_{\sigma_n,\sigma_{n'}}-D_{n',\sigma_{n'};n,\sigma_n}$.
$H$ is the semiconductor hamiltonian and $B_{n,\sigma_n}^\dag$ is
the creation operator of the  exciton $n=(\nu_n,\mathbf{Q}_n)$,
with energy
$E_n=\epsilon_{
\nu_n}+\hbar^2\mathbf{Q}_n^2/2(m_e+\nolinebreak m_h)$ and
electron spin
$\sigma_n$. (The hole ``spin'' playing no r\^{o}le here, we drop
it to simplify the notations). $B_{n,\sigma_n}^\dag$ is linked
to the electron and hole creation operators $a_{\mathbf{k},s}
^\dag$ and $b_\mathbf{k}^\dag$ by 
\begin{equation}
B_{n,\sigma_n}^\dag=\sum_{\mathbf{p}}\langle
\mathbf{p}|x_{\nu_n}\rangle\,
b_{-\mathbf{p}+\alpha_h\mathbf{Q}_n}^\dag\,
a_{\mathbf{p}+\alpha_e\mathbf{Q}_n,\sigma_n}^\dag\ ,
\end{equation}
\begin{equation}
b_{\mathbf{k}_h}^\dag\,a_{\mathbf{k}_e,s}^\dag=\sum_\nu
\langle x_\nu|\alpha_h\mathbf{k}_e-\alpha_e\mathbf{k}_h\rangle\,
B_{\nu,\mathbf{k}_e+\mathbf{k}_h,s}^\dag\ ,
\end{equation}
where $\alpha_{e,h}=m_{e,h}/(m_e+m_h)$. By using the explicit
expression of $V_{n,\sigma_n}^\dag$ deduced from its definition
$^{(8)}$, eq.\ (1) leads to
\begin{equation}
\Xi_{n'\mathbf{k}'n\mathbf{k}}^{\mathrm{dir}}
=\delta_{\mathbf{Q}_{n'}+\mathbf{k}',\mathbf{Q}_n
+\mathbf{k}}\,W_{\nu_{n'}\nu_n}(\mathbf{Q}_{n'}-\mathbf{Q}_n)
\ ,
\end{equation}
\begin{equation}
W_{\nu'\nu}(\mathbf{q})=V_{\mathbf{q}}\,\langle x_{\nu'}
|e^{i\alpha_h\mathbf{q}.\mathbf{r}}-e^{-i
\alpha_e\mathbf{q}.\mathbf{r}}|x_\nu\rangle\ ,
\end{equation}
$V_{\mathbf{q}}$ being the Fourier transform of the
Coulomb potential. Let us mention that 
\begin{equation}
\Xi_{n'\mathbf{k}'n\mathbf{k}}^{\mathrm{dir}}=
\int
d\mathbf{r}_e\,d\mathbf{r}_{e'}\,d\mathbf{r}_h
\phi_{n'}^\ast(\mathbf{r}_e,\mathbf{r}_h)\,
f_{\mathbf{k'}}^\ast(\mathbf{r}_{e'})\,\left[
\frac{e^2}{|\mathbf{r}_{e'}
-\mathbf{r}_e|}-\frac{e^2}{|\mathbf{r}_{e'}-\mathbf{r}_h|}\right]\,
\phi_n(\mathbf{r}_e,\mathbf{r}_h)\,
f_{\mathbf{k}}(\mathbf{r}_{e'})\ ,
\end{equation}
where $f_{\mathbf{k}}(\mathbf{r})=e^{i\mathbf{k}.
\mathbf{r}}/\sqrt{\mathcal{V}}$ is the
free-particle wave function
while \linebreak
$\phi_n(\mathbf{r}_e,\mathbf{r}_h)=\langle\mathbf{r}_{eh}
|x_{\nu_n}\rangle\,f_{\mathbf{Q}_n}(\mathbf{R}_{eh})$, with
$\mathbf{r}_{eh}=\mathbf{r}_e-\mathbf{r}_h$ and $\mathbf{R}_{eh}
=(m_e\mathbf{r}_e+m_h\mathbf{r}_h)/(m_e+m_h)$, is 
the total wave function of the $n$ exciton :
 $\Xi_{n'\mathbf{k}'n\mathbf{k}}^{\mathrm{dir}}$ thus
corresponds to the scattering of a ($n$,
$\mathbf{k}$) state into a
($n'$, $\mathbf{k}'$) state
induced by Coulomb interactions \emph{between} the exciton and
the electron, the $n$ and $n'$ excitons
being made with the \emph{same} electron-hole
pair $(e,h)$.

If we turn to
$\Lambda_{n'\mathbf{k}'n\mathbf{k}}$, eqs.\ (2,3) and the
explicit expression of $D_{n',\sigma';n,\sigma}$ deduced from its
definition $^{(8)}$ lead to 
\begin{equation}
\Lambda_{n'\mathbf{k}'n\mathbf{k}}=\delta_{\mathbf{Q}_{n'}+
\mathbf{k}',\mathbf{Q}_n+\mathbf{k}}\,L_{\nu_{n'},
\beta_x\mathbf{k'}
-\beta_e\mathbf{Q}_{n'};\nu_n,\beta_x\mathbf{k}-
\beta_e\mathbf{Q}_n}\ ,
\end{equation}
\begin{equation}
L_{\nu'\mathbf{p'}\nu\mathbf{p}}
=\langle x_{\nu'}|\mathbf{p}+
\alpha_e\mathbf{p'}\rangle\,\langle\mathbf{p'}+\alpha_e
\mathbf{p}|x_\nu\rangle\ ,
\end{equation}
with $\beta_e=1-\beta_x=m_e/(2m_e+m_h)$.
We can mention that
\begin{equation}
\Lambda_{n'\mathbf{k}'n\mathbf{k}}=\int
d\mathbf{r}_e\,d\mathbf{r}_{e'}\, d\mathbf{r}_h\,\phi_{n'}^\ast
(\mathbf{r}_e,\mathbf{r}_h)\,
f_{\mathbf{k'}}^\ast(\mathbf{r}_{e'})\,
\phi_n(\mathbf{r}_{e'},\mathbf{r}_h)\,
f_{\mathbf{k}}(\mathbf{r}_e)\ ,
\end{equation}
which clearly shows that the
$(n,\mathbf{k})$ and $(n',\mathbf{k}')$ states are coupled
by $\Lambda_{n'\mathbf{k'}n\mathbf{k}}$ due to their electron
exchange independently from any Coulomb process. This possible
exchange also leads to
\begin{equation}
B_{n,\sigma}^\dag\,a_{\mathbf{k},s}^\dag=
-\sum_{n',\mathbf{k}'}
\Lambda_{n'\mathbf{k}'n\mathbf{k}}\,
B_{n',s}^\dag\,a_{\mathbf{k'},\sigma}^\dag
\ ,
\end{equation}
while two exchanges reduce to identity :
\begin{equation}
\sum_{n'',\mathbf{k''}}\Lambda_{n'\mathbf{k'}
n''\mathbf{k''}}\,\Lambda_{n''\mathbf{k''}
n\mathbf{k}}=\delta_{nn'}\,\delta_{\mathbf{kk'}}\
.
\end{equation}

\section{Trion creation operators}

The $\mathrm{X}^-$ trions being made of two electrons and one
hole, their creation operators write in terms
of
$b_{\mathbf{k}_h}^\dag a_{\mathbf{k}_e}^\dag a_{\mathbf{k}
_{e'}}^\dag$. According to eq.\ (4), they can also be written in
terms of $B_{n}^\dag a_{\mathbf{k}}^\dag$, with
$\mathbf{Q}_n +\mathbf{k}=\mathbf{K}_i$, $\mathbf{K}_i$ being
the trion total momentum. 

Let us consider the operators
\begin{equation}
\mathrm{T}_{i;S,S_z}^\dag=\sum_{\nu,\mathbf{p}}\tilde{\psi}_{\nu,
\mathbf{p}}^{(\eta_i;S)}\,\mathcal{T}_{\nu,\mathbf{p},\mathbf{K}_i;S,S_z}
^\dag\ ,
\end{equation}
where $i$ stands for $(\eta_i,\mathbf{K}_i)$, the four
free exciton-electron creation operators 
$\mathcal{T}_{\nu,\mathbf{p},\mathbf{K};S,S_z}^\dag$ being given
by
\begin{eqnarray}
\mathcal{T}_{\nu,\mathbf{p},\mathbf{K};1,\pm1}^\dag &=
&B_{\nu,-\mathbf{p}+\beta_x\mathbf{K},\pm}^\dag\,a_
{\mathbf{p}+\beta_e\mathbf{K},\pm}^\dag\ ,\nonumber
\\
\mathcal{T}
_{\nu,\mathbf{p},\mathbf{K};S,0}^\dag 
&=&
\left(B_{\nu,-\mathbf{p}+\beta_x\mathbf{K},+}^\dag\,
a_{\mathbf{p}+\beta_e\mathbf{K},-}^\dag-(-1)^S
B_{\nu,-\mathbf{p}+\beta_x\mathbf{K},-}^\dag\,
a_{\mathbf{p}+\beta_e\mathbf{K},+}
^\dag\right)/\sqrt{2}\ .
\end{eqnarray}
Due to eqs.\ (11,3,4,9), these operators are such that
\begin{equation}
\mathcal{T}_{\nu,\mathbf{p},\mathbf{K};S,S_z}^\dag
=(-1)^S\sum_{\nu',
\mathbf{p'}}L_{\nu'\mathbf{p'}\nu\mathbf{p}}
\mathcal{T}_{\nu',\mathbf{p'},\mathbf{K};S,S_z}^\dag\ ,
\end{equation}
\begin{equation}
\langle
v|\mathcal{T}_{\nu',\mathbf{p'},\mathbf{K'};S',S'_z}\,\mathcal{T}
_{\nu,\mathbf{p},\mathbf{K};S,S_z}^\dag
|v\rangle=\delta_{\mathbf{K'},\mathbf{K}}\,\delta_{S',S}\,
\delta_{S'_z,S_z}\left(\delta_{\nu',\nu}\,\delta_{\mathbf{p'},
\mathbf{p}}
+(-1)^S L_{\nu'\mathbf{p'}\nu\mathbf{p}}\right)\ .
\end{equation}

Equation (15) allows to replace the trion prefactors
$\tilde{\psi}$ in eq.\ (13) by $\psi$ defined as
\begin{equation}
\psi_{\nu,\mathbf{p}}^{(\eta_i;S)}=\frac{1}{2}
\left(\tilde{\psi}_{\nu,\mathbf{p}}^{(\eta_i;S)}
+(-1)^S\sum_{\nu',
\mathbf{p'}}L_{\nu\mathbf{p}\nu'\mathbf{p'}}
\tilde{\psi}_{\nu',\mathbf{p'}}^{(\eta_i;S)}\right)\ ,
\end{equation}
so that, due to eq.\ (12), these prefactors now verify
\begin{equation}
\psi_{\nu,\mathbf{p}}^{(\eta_i;S)}=(-1)^S\sum_{\nu',
\mathbf{p'}}
L_{\nu\mathbf{p}\nu'\mathbf{p'}}\,
\psi_{\nu',\mathbf{p'}}^{(\eta_i;S)}\ ,
\end{equation}
which just states that they stay invariant under the possible
exchange corresponding to eq.\ (11). From eq.\ (13), with
$\tilde{\psi}$ replaced by $\psi$, and eqs.\ (16,18), one can
easily check that
\begin{equation}
\langle v|\mathcal{T}_{\nu,\mathbf{p},\mathbf{K};S,S_z}\,
\mathrm{T}_{\eta_i,\mathbf{K}_i;S_i,S_{iz}}|v\rangle=2\,
\delta_{S,S_i}\,\delta_{S_z,S_{iz}}\,\delta_{\mathbf{K},
\mathbf{K}_i}\,\psi_{\nu,\mathbf{p}}^{(\eta_i;S)}\ .
\end{equation}

$\mathrm{T}_{i;S,S_z}^\dag|v\rangle$ is indeed a trion,
i.\ e.\ an eigenstate of
$H$, with the energy $\mathcal{E}_{i;S}$, if
\begin{equation}
\langle
v|\mathcal{T}_{\nu,\mathbf{p},\mathbf{K}_i;S,S_z}\,(H-\mathcal{E}_{i;S})\,
\mathrm{T}_{i;S,S_z}^\dag|v\rangle=0.
\end{equation}
By using eqs.\ (14,1,5,19), we find that eq.\ (20) leads to
\begin{equation}
(\epsilon_\nu+\frac{\hbar^2\mathbf{p}^2}{2\mu_t}-\varepsilon
_{\eta_i;S})
\psi_{\nu,\mathbf{p}}^{(\eta_i;S)}+\sum_{\nu',\mathbf{p'}}
W_{\nu\nu'}(-\mathbf{p}+\mathbf{p'})
\psi_{\nu',\mathbf{p'}}^{(\eta_i;S)}\ ,
\end{equation}
where we have set $\mathcal{E}_{i;S}=\varepsilon_{\eta_i;S}
+\hbar^2\mathbf{K}_i^2/2(2m_e+m_h)$,
$\mu_t$ being the exciton-electron relative motion mass, 
$\mu_t^{-1}=m_e^{-1}+(m_e+m_h)^{-1}$.

It can be surprising to note that the integral equation (21)
verified by the trion prefactors $\psi_{\nu,\mathbf{p}}^{
(\eta_i;S)}$ only depends 
on the direct Coulomb scattering
$\Xi_{n\mathbf{k}n'\mathbf{k'}}^{\mathrm{dir}}$, through
$W_{\nu\nu'}(-\mathbf{p}+\mathbf{p'})$. Exchange couplings
$\Lambda_{n\mathbf{k}n'\mathbf{k'}}$, through $L_{\nu\mathbf{p}
\nu'\mathbf{p'}}$, do not appear in it. They
are actually hidden in the invariance relation (18) between the
$\psi_{\nu,\mathbf{p}}^{(\eta_i;S)}$'s.

The trion orbital wave function deduced
from eqs.\ (13,14) reads
\begin{eqnarray}
F_{i;S}(\mathbf{r}_e,\mathbf{r}_{e'},\mathbf{r}_h)&=&
f_{\mathbf{K}_i}(\mathbf{R}_{ee'h})\sum_{\nu,\mathbf{p}}
\psi_{\nu,\mathbf{p}}^{(\eta_i;S)}\left[\langle\mathbf{r}_{eh}
|x_\nu\rangle\,f_\mathbf{p}(\mathbf{u}_{e'eh})+(-1)^S
\langle\mathbf{r}_{e'h}
|x_\nu\rangle\,f_\mathbf{p}(\mathbf{u}_{ee'h})\right]/\sqrt{2}
\nonumber \\ &\equiv&
\sqrt{2}\,
f_{\mathbf{K}_i}(\mathbf{R}_{ee'h})\sum_{\nu,\mathbf{p}}
\psi_{\nu,\mathbf{p}}^{(\eta_i;S)}\,\langle\mathbf{r}_{eh}
|x_\nu\rangle\,f_\mathbf{p}(\mathbf{u}_{e'eh})\ ,
\end{eqnarray}
with
$\mathbf{R}_{ee'h}=(m_e\mathbf{r}_e+m_{e'}\mathbf{r}_{e'}+m_h
\mathbf{r}_h)/(2m_e+m_h)$ being the trion center of mass
position, and
$\mathbf{u}_{e'eh}=\mathbf{r}_{e'}-\mathbf{R}_{eh}$ being the
distance between the electron $e'$ and the center of mass of the
exciton made with $(e,h)$, the two terms of the first line of
eq.\ (22) being equal due to eqs.\ (18,9). 
As a consequence, the $\psi$'s can be
obtained from the trion orbital wave function through
\begin{equation}
\sqrt{2}\psi_{\nu,\mathbf{p}}^{(\eta_i;S)}=\int
d\mathbf{R}\,d\mathbf{r}
\,d\mathbf{u}\,f_{\mathbf{K}_i}^\ast(\mathbf{R})\,
\langle x_\nu|\mathbf{r}\rangle\,f_{\mathbf{p}}^\ast(\mathbf{u})
\,F_{i;S}(\mathbf{R}+\alpha_h\mathbf{r}-\beta_e\mathbf{u},\,
\mathbf{R}+\beta_x\mathbf{u},\,\mathbf{R}-\alpha_e\mathbf{r}
-\beta_e\mathbf{u})\ .
\end{equation}

\section{Trion ladder diagrams}

It is known $^{(9)}$ that excitons correspond to ``ladder
diagrams'' between one free electron and one free hole, which
originate from the electron-hole Coulomb potential V. By writing
the semiconductor hamiltonian as $H=H_0+V$, these
diagrams simply result from the iteration of
$(a-H)^{-1}=(a-H_0)^{-1}+(a-H)^{-1}\,V\,(a-H_0)^{-1}$
acting on one free electron-hole pair.

For trions, the problem appears at first as much more tricky, due
to the composite nature of the exciton. There is indeed no way
to write $H$ as $H_0'+V'$, with $V'$ being an
exciton-electron potential : This is in fact \emph{the major
difficulty} of all problems dealing with interacting excitons.

Our ``commutation technique'' allows to
overcome this difficulty. Indeed, by using the equation which
defines the ``creation potential'' $V_{n,\sigma_n}^\dag$, we can
show that 
\begin{equation}
\frac{1}{a-H}\,B_{n,\sigma_n}^\dag=B_{n,\sigma_n}^\dag
\,\frac{1}{a-H-E_n}+\frac{1}{a-H}\,
V_{n,\sigma_n}^\dag\,\frac{1}{a-H-E_n}\ .
\end{equation}
For $a=\Omega+i\eta$, this equation, along with
eqs.\ (1,5), gives
$(\Omega-H+i\eta)^{-1}$ acting on one exciton-free electron pair
as
\begin{eqnarray}
\frac{1}{\Omega-H+i\eta}\,\mathcal{T}_{\nu,\mathbf{p},\mathbf{K};
S,S_z}^\dag|v\rangle=\frac{1}{\Omega-E_{\nu,\mathbf{p},\mathbf{K}}+i\eta}
\,\left[\mathcal{T}_{\nu,\mathbf{p},\mathbf{K};S,S_z}^\dag|v
\rangle\right.\hspace{3cm}\nonumber
\\ \left.+\sum_{\nu',\mathbf{p'}}W_{\nu'\nu}(-\mathbf{p'}+
\mathbf{p})\,\frac{1}{\Omega-H+i\eta}\,
\mathcal{T}_{\nu',\mathbf{p'},\mathbf{K};S,S_z}
^\dag|v\rangle\right]\ ,
\end{eqnarray}
with
$E_{\nu,\mathbf{p},\mathbf{K}}=\epsilon_\nu+\hbar^2\mathbf{p}^2
/2\mu_t+\hbar^2\mathbf{K}^2/2(2m_e+m_h)$.

The iteration of the above equation leads to
\begin{equation}
\frac{1}{\Omega-H+i\eta}\,\mathcal{T}_{\nu,\mathbf{p},\mathbf{K}
;S,S_z}^\dag| v\rangle=\sum_{\nu',\mathbf{p'}}A_{\nu'\mathbf{p'}
\nu\mathbf{p}}(\Omega,\mathbf{K})\,\mathcal{T}_{\nu',\mathbf{p'},
\mathbf{K};S,S_z}|v\rangle\ ,
\end{equation}
in which we have set
\begin{eqnarray}
A_{\nu'\mathbf{p'}\nu\mathbf{p}}(\Omega,\mathbf{K})=\left[
\delta_{\nu',\nu}\,\delta_{\mathbf{p'},\mathbf{p}}+\frac{1}
{\Omega-E_{\nu',\mathbf{p'},\mathbf{K}}+i\eta}\left(
W_{\nu'\nu}(-\mathbf{p'}+\mathbf{p})\right.\right.\hspace{4cm}
\nonumber
\\
\left.\left.+\sum_{\nu_1,\mathbf{p}_1}
\frac{W_{\nu'\nu_1}(-\mathbf{p'}+\mathbf{p}_1)\,W_{\nu_1\nu}
(-\mathbf{p}_1+\mathbf{p})}{\Omega-E_{\nu_1,\mathbf{p}_1,
\mathbf{K}}+i\eta}+\cdots\right)\right]\frac{1}{\Omega
-E_{\nu,\mathbf{p},\mathbf{K}}+i\eta}\ .
\end{eqnarray}
This $A_{\nu'\mathbf{p'}\nu\mathbf{p}}(\Omega,\mathbf{K})$ can
be formally rewritten as
\begin{equation}
A_{\nu'\mathbf{p'}\nu\mathbf{p}}(\Omega,\mathbf{K})=
\left[\delta_{\nu',
\nu}\,\delta_{\mathbf{p'},\mathbf{p}}+\frac{\tilde{W}_{\nu'
\mathbf{p'}\nu\mathbf{p}}(\Omega,\mathbf{K})}
{\Omega-E_{\nu',\mathbf{p'},\mathbf{K}}+i\eta}\right]
\frac{1}
{\Omega-E_{\nu,\mathbf{p},\mathbf{K}}+i\eta}\ ,
\end{equation}
where the ``renormalized exciton-electron interaction''
$\tilde{W}_{\nu'
\mathbf{p'}\nu\mathbf{p}}(\Omega,\mathbf{K})$ verifies
\begin{equation}
\tilde{W}_{\nu'\mathbf{p'}\nu\mathbf{p}}(\Omega,\mathbf{K})=
W_{\nu'\nu}(-\mathbf{p'}+\mathbf{p})+\sum_{\nu_1,\mathbf{p}_1}
\frac{\tilde{W}_{\nu'\mathbf{p'}\nu_1\mathbf{p}_1}(\Omega,
\mathbf{K})\,W_{\nu_1\nu}(-\mathbf{p}_1+\mathbf{p})}
{\Omega-E_{\nu_1,\mathbf{p}_1,\mathbf{K}}+i\eta}\ .
\end{equation}
This iteration is
shown in fig.\ (1). Before going further, let us note that
\begin{equation}
\frac{1}{\Omega-E_{\nu_1,\mathbf{p}_1,\mathbf{K}}+i\eta}\equiv
\int\frac{id\omega_1}{2\pi}g_e(\Omega+\omega_1,\mathbf{p}_1
+\beta_e
\mathbf{K})\,g_x(-\omega_1;\nu_1,-\mathbf{p}_1+\beta_x\mathbf{K})=
G_{xe}(\Omega;\nu_1,\mathbf{p}_1,\mathbf{K})\ ,
\end{equation}
where $g_e(\omega,\mathbf{k})=(\omega-\hbar^2\mathbf{k}^2/2m_e
+i\eta)^{-1}$ is the usual free electron Green's function for an
empty Fermi sea, while $g_x(\omega;n)=(\omega-E_n+i\eta)^{-1}$
is the free boson-exciton Green's function, as if the
excitons were non-interacting bosons, i.\ e.\ if all the $\Xi$'s
\emph{and} $\Lambda$'s were set equal to zero. $G_{xe}$ can be
seen as the propagator of a free exciton-electron pair. It is
quite similar to the free electron-hole pair propagator $G_{eh}$
appearing in exciton diagrams (see eq.\ 2.10 of ref.\ (9)).

The simplest way to obtain 
$A_{\nu'\mathbf{p'}\nu\mathbf{p}}(\Omega,\mathbf{K})$ is
to insert the trion closure relation between the two operators
of the l.h.s.\ of eq.\ (26) and to project this equation over 
$\langle v|\mathcal{T}_{\nu'',\mathbf{p''},\mathbf{K};S,S_z}$.
By using eqs.\ (19,16), we obtain
\begin{equation}
\sum_{\eta_i}\frac{4\,\psi_{\nu'',\mathbf{p''}}^{(\eta_i;S)}\,
\left(\psi_{\nu,\mathbf{p}}^{(\eta_i;S)}\right)^\ast}
{\Omega-\mathcal{E}_{\eta_i,\mathbf{K};S}+i\eta}=
A_{\nu''\mathbf{p''}\nu\mathbf{p}}(\Omega,\mathbf{K})
+(-1)^S\sum_{\nu',\mathbf{p'}}
L_{\nu''\mathbf{p''}\nu'\mathbf{p'}}\,A_{\nu'\mathbf{p'}\nu
\mathbf{p}}(\Omega,\mathbf{K})\ .
\end{equation}
$A_{\nu''\mathbf{p''}\nu\mathbf{p}}(\Omega,\mathbf{K})$ is then
obtained by adding the above equations for
$S=0$ and
$S=1$. Using eq.\ (28), we thus get the sum of the
exciton-electron ladder ``rungs'' as
\begin{equation}
\tilde{W}_{\nu'\mathbf{p'}\nu\mathbf{p}}(\Omega,\mathbf{K})=
\left[-\delta_{\nu',\nu}\,\delta_{\mathbf{p'},\mathbf{p}}+
\frac{1}{G_{xe}(\Omega;\nu',\mathbf{p'},\mathbf{K})}
\sum_{\eta_i,S}\frac{2\,\psi_{\nu',\mathbf{p'}}^{(\eta_i;S)}
\left(\psi_{\nu,\mathbf{p}}^{(\eta_i;S)}\right)^\ast}
{\Omega-\mathcal{E}_{\eta_i,\mathbf{K};S}+i\eta}\right]
\frac{1}{G_{xe}(\Omega;\nu,\mathbf{p},\mathbf{K})}\ .
\end{equation}
This result is quite similar to the ``renormalized electron-hole
Coulomb interaction" appearing in electron-hole ladder diagrams,
as given for example in eq.\ (2.18) of ref.\ (9).

\section{Conclusion}

We have generated the exciton-electron ladder diagrams
associated to trions. They will allow to treat trions in a
similar way as excitons, with respect to many-body effects in
which they can be involved. This work relies on a new
commutation technique for excitons interacting with electrons
which takes exactly into account the possible exchange between
carriers.

\newpage
\hbox to \hsize{\hfill REFERENCES \hfill}
\vspace{1cm}

\noindent
(1) E. HYLLERAAS, Phys.\ Rev.\ \underline{71}, 491 (1947).

\noindent
(2) M. LAMPERT, Phys.\ Rev.\ Lett.\ \underline{1}, 450 (1958).

\noindent
(3) G. FINKELSTEIN, H.SHTRIKMAN, I. BAR-JOSEPH, Phys.\ Rev.\
Lett.\ \underline{74}, 976 (1995).

\noindent
(4) A.J. SHIELDS, M. PEPPER, D.A. RITCHIE, M.Y. SIMMONS, G.A.
JONES, Phys.\ Rev.\ B \underline{51}, 18049 (1995).

\noindent
(5) S.A. BROWN, J.F. YOUNG, J.A. BRUM, P. HAWRYLAK, Z.
WASILEWSKI, Phys.\ Rev.\ B, Rapid Com. \underline{54}, 11082
(1996).

\noindent
(6) R. KAUR, A.J. SHIELDS, J.L. OSBORNE, M.Y. SIMMONS, D.A.
RITCHIE, M. PEPPER, Phys.\ Stat.\ Sol.\ \underline{178}, 465
(2000).

\noindent
(7) V. HUARD, R.T. COX, K. SAMINADAYAR, A. ARNOULT, S. TATARENKO,
Phys.\ Rev.\ Lett.\ \underline{84}, 187 (2000).

\noindent
(8) M. COMBESCOT, O. BETBEDER-MATIBET, Europhys.\ Lett.\ 
\underline{58}, 87 (2002) ; O. BETBEDER-MATIBET, M. COMBESCOT,
Eur.\ Phys.\ J.\ B \underline{27}, 505 (2002).

\noindent
(9) O. BETBEDER-MATIBET, M. COMBESCOT, Eur.\ Phys.\ J.\ B
\underline{22}, 17 (2001).

\newpage
\hbox to \hsize{\hfill FIGURE CAPTIONS \hfill}
\vspace{1cm}

\noindent
Fig (1)

(a) Direct Coulomb scattering between one ``free'' exciton and
one free electron.

(b) Renormalized free exciton-electron interaction as given by
the integral equation (29). It corresponds to the sums of one,
two,\ldots ladder ``rungs'' between one ``free'' exciton and one
free electron.

\end{document}